\documentclass[aps,prl,reprint,superscriptaddress]{revtex4-1}

\usepackage{graphicx}% Include figure files

\usepackage[colorlinks=true,citecolor=blue,linkcolor=blue,pdfstartview=FitH,pdfauthor={Goennenwein}]{hyperref}
\usepackage{color}
\begin{document}

%\preprint{}
\title{Non-local magnetoresistance in YIG/Pt nanostructures}

% Please use the appropriate macro for reach each type of information

\author{Sebastian T. B. Goennenwein}
\email[]{goennenwein@wmi.badw.de}
%\thanks{}
\affiliation{Walther-Mei{\ss}ner-Institut, Bayerische Akademie der Wissenschaften, Garching, Germany}
\affiliation{Nanosystems Initiative Munich (NIM), Schellingstra{\ss}e 4, M\"{u}nchen, Germany}
\affiliation{Physik-Department, Technische Universit\"{a}t M\"{u}nchen, Garching, Germany}
\author{Richard Schlitz}
\affiliation{Walther-Mei{\ss}ner-Institut, Bayerische Akademie der Wissenschaften, Garching, Germany}
\affiliation{Physik-Department, Technische Universit\"{a}t M\"{u}nchen, Garching, Germany}
\author{Matthias Pernpeintner}
\affiliation{Walther-Mei{\ss}ner-Institut, Bayerische Akademie der Wissenschaften, Garching, Germany}
\affiliation{Nanosystems Initiative Munich (NIM), Schellingstra{\ss}e 4, M\"{u}nchen, Germany}
\affiliation{Physik-Department, Technische Universit\"{a}t M\"{u}nchen, Garching, Germany}
\author{Matthias Althammer}
\affiliation{Walther-Mei{\ss}ner-Institut, Bayerische Akademie der Wissenschaften, Garching, Germany}
\author{Rudolf Gross}
\affiliation{Walther-Mei{\ss}ner-Institut, Bayerische Akademie der Wissenschaften, Garching, Germany}
\affiliation{Nanosystems Initiative Munich (NIM), Schellingstra{\ss}e 4, M\"{u}nchen, Germany}
\affiliation{Physik-Department, Technische Universit\"{a}t M\"{u}nchen, Garching, Germany}
\author{Hans Huebl}
\affiliation{Walther-Mei{\ss}ner-Institut, Bayerische Akademie der Wissenschaften, Garching, Germany}
\affiliation{Nanosystems Initiative Munich (NIM), Schellingstra{\ss}e 4, M\"{u}nchen, Germany}
\affiliation{Physik-Department, Technische Universit\"{a}t M\"{u}nchen, Garching, Germany}

\date{\today}

\begin{abstract}
We study the local and non-local magnetoresistance of thin Pt strips deposited onto yttrium iron garnet. The local magnetoresistive response, inferred from the voltage drop measured along one given Pt strip upon current-biasing it, shows the characteristic magnetization orientation dependence of the spin Hall magnetoresistance. We simultaneously also record the non-local voltage appearing along a second, electrically isolated, Pt strip, separated from the current carrying one by a gap of a few 100\,nm. The corresponding non-local magnetoresistance exhibits the symmetry expected for a magnon spin accumulation-driven process, confirming the results recently put forward by Cornelissen et al.\cite{Magnon-MR:Cornelissen:arXiv2015}. Our magnetotransport data, taken at a series of different temperatures as a function of magnetic field orientation, rotating the externally applied field in three mutually orthogonal planes, show that the mechanisms behind the spin Hall and the non-local magnetoresistance are qualitatively different. In particular, the non-local magnetoresistance vanishes at liquid Helium temperatures, while the spin Hall magnetoresistance prevails.
\end{abstract}

\pacs{85.75.-d,72.25.Mk,75.75.Cd,75.50.Dd}
%73.40.Rw	Metal-insulator-metal structures
%75.50.Dd	Nonmetallic ferromagnetic materials
%75.75.Cd	Fabrication of magnetic nanostructures
%72.25.Mk	Spin transport through interfaces
%72.25.Ba	Spin polarized transport in metals
%85.75.-d	Magnetoelectronics; spintronics: devices exploiting spin polarized transport or integrated magnetic fields
%\keywords{}

\maketitle

\begin{figure}
 \includegraphics[width=8.5cm]{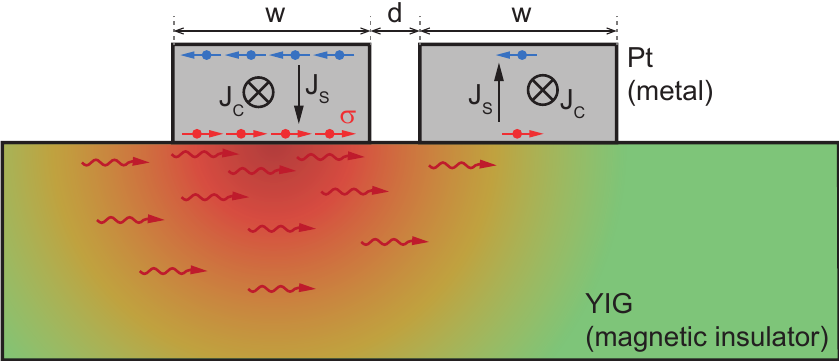}%
 \caption{\label{Fig:model} Sketch of the magnon-mediated magnetoresistance (MMR) following Cornelissen et al. \cite{Magnon-MR:Cornelissen:arXiv2015}. Driving a charge current $\mathbf{J}_{c}$ through the left strip of platinum (Pt) results in an orthogonal spin current with propagation direction $\mathbf{J}_{s}$ and spin polarization $\mathbf{\sigma}$. The spin accumulation in the metal induces a magnon accumulation (wiggly red arrows) in the adjacent magnetic insulator yttrium iron garnet (YIG). This magnon accumulation decays with increasing distance to the current-carrying Pt injector strip (shaded region). If a second Pt strip, electrically isolated from the first, is within the range of magnon accumulation, a spin current will flow back from the magnetic insulator into the second Pt strip and give rise to an inverse spin Hall charge current there. The two Pt strips of width $w$ are separated by an edge-to-edge distance $d$.}
 \end{figure}

% ---------------------------------- text
Magneto-resistive phenomena are powerful probes for the magnetic properties. The anisotropic magnetoresistance in ferromagnetic metals \cite{McGuire:MTr-AMR:IEEETM:1975}, or the giant magnetoresistance \cite{Baibich:Fert:GMR:1988} and the tunneling magnetoresistance \cite{TMR:Moodera:PRL:1995} observed in thin film heterostructures based on magnetic metals are widely used in sensing and data storage applications \cite{Ohandley:ModernMagneticMaterials}. Heterostructures consisting of an insulating magnetic layer (such as yttrium iron garnet $\mathrm{Y}_{3}\mathrm{Fe}_{5}\mathrm{O}_{12}$ (YIG)) and a heavy metal (such as platinum (Pt)) also exhibit a magnetoresistance \cite{Nakayama:SMR:PRL:2013,Althammer:SMR:experiment:PRB:2013,SMR:Vlietstra:PRB:2013,SMR:Hahn:PRB:2013}. This so-called spin Hall magnetoresistance (SMR) is due to spin torque transfer across the magnetic insulator/metal interface \cite{Chen:SMR:theory:PRB:2013}. Owing to the spin Hall effect \cite{SHE:Dyakonov:Perel:JETPLett:1971,SHE:Hirsch:PRL:1999}, a spin accumulation $\mathbf{\sigma}$ arises in the metal, at the interface to the magnet (cf.~Fig.\,\ref{Fig:model}). Given that $\mathbf{\sigma}$ is not collinear with the magnetization $\mathbf{M}$ in the magnet, the spin accumulation can exert a torque proportional to $\mathbf{M}\times (\mathbf{M} \times \mathbf{\sigma})$ on $\mathbf{M}$. In other words, a finite spin current flow across the interface is possible if $\mathbf{M}$ and $\mathbf{\sigma}$ enclose a finite angle. Since the spin current flow across the interface represents a dissipation channel for the charge transport in the metal layer, its resistance therefore will change with the magnetization orientation as $\Delta R\propto \mathbf{M}\times (\mathbf{M} \times \mathbf{\sigma})$ \cite{Nakayama:SMR:PRL:2013,Althammer:SMR:experiment:PRB:2013,SMR:Vlietstra:PRB:2013,SMR:Hahn:PRB:2013}. In order to experimentally resolve this SMR fingerprint, magneto-resistance measurements as a function of magnetization orientation in at least three different planes are mandatory \cite{Althammer:SMR:experiment:PRB:2013}.

Recently, Cornelissen et al.~\cite{Magnon-MR:Cornelissen:arXiv2015} discovered a non-local magneto-resistance effect in YIG/Pt heterostructures and attributed it to magnon accumulation and transport. We will refer to this effect as magnon-mediated magneto-resistance (MMR) in the following. The MMR is observed in two parallel Pt strips separated by a distance $d$ deposited onto YIG, as sketched in Fig.\,\ref{Fig:model}. Driving a charge current through the left Pt strip will generate a spin accumulation $\mathbf{\sigma}$ in Pt at the interfaces. As shown in Fig.\,\ref{Fig:model}, $\mathbf{\sigma}$ is perpendicular to the direction of charge current flow $\mathbf{J}_{c}$ and orthogonal to the spin current flow $\mathbf{J}_{s}$ across the interface. This spin accumulation in particular also induces a magnon (spin) accumulation in YIG \cite{Magnon-MR:Cornelissen:arXiv2015} -- an effect which usually is assumed small and ignored in the treatment of SMR \cite{Chen:SMR:theory:PRB:2013}. The non-equilibrium magnon accumulation diffuses out into the magnetic insulator, as indicated by the shading in Fig.\,\ref{Fig:model}. Given that the second Pt electrode is close enough such that the diffusing magnons can reach it before decaying, the magnon accumulation will drive a spin current back into the second Pt electrode. In turn, this spin current then generates an inverse spin Hall charge current in the second Pt electrode. A large non-local charge current is expected for $\mathbf{M} || \mathbf{\sigma}$, since the magnons (the spin angular momenta) beneath the first Pt strip then can diffuse across the gap to the second Pt strip. For $\mathbf{M} \perp \mathbf{\sigma}$, in contrast, the non-local signal should be significantly reduced, since now spin torque transfer suppresses the magnon accumulation and/or the magnon propagation. This picture of a non-local magnon-based magnetoresistance, put forward by Cornelissen et al. in Ref.~\cite{Magnon-MR:Cornelissen:arXiv2015}, to date only has been tested against non-local inverse spin Hall voltage data taken as a function of magnetic field orientation for the magnetic field in the plane of the YIG film. Neither a direct comparison of the non-local magnetoresistance with the SMR, nor a study of the evolution of the non-local voltage as a function of out-of-plane magnetization orientation, have been put forward. Note also that the MMR is different from the non-local electrical detection of spin pumping \cite{Costache:vanWees:spin-pumping:experiment:PRL2006,spin-pumping-HL:Shiraishi:PRL:2013} or magnetoresistance experiments in metallic spin valves \cite{Jedema:2001,Tang:in:Awschalom-Halbleiter-Spin-Buch}, since the YIG (the magnetically ordered material) only passively acts as a 'spin transport' medium, contacted with conventional metallic nano-electrodes.

\begin{figure}
 \includegraphics[width=8.5cm]{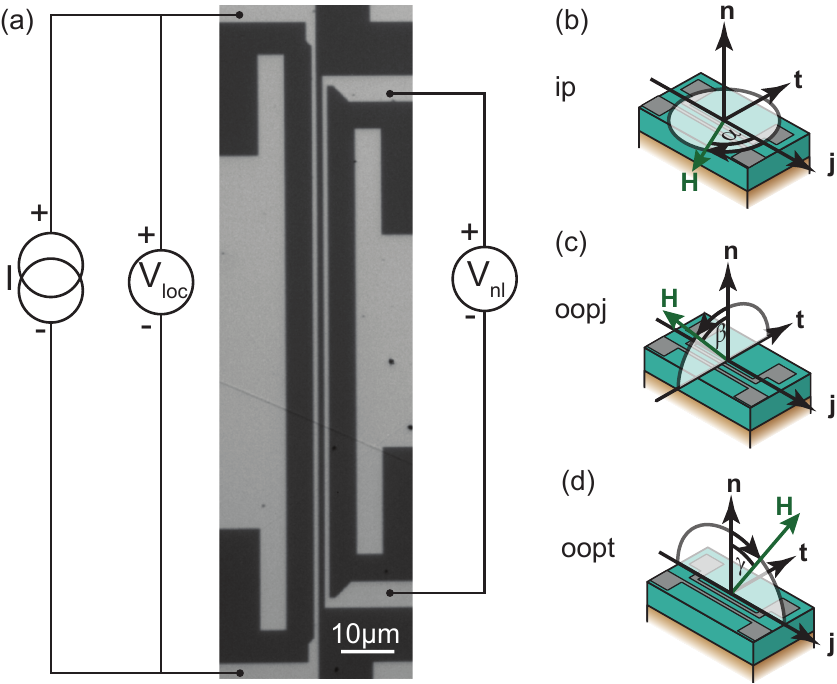}%
 \caption{\label{Fig:expt-schematics} (a) Optical micrograph of a typical YIG/Pt nanostructure. The two bright thin vertical lines in the center of the figure are the two Pt strips under investigation, the YIG film beneath appears black. %The thin black lines show the electrical measurement scheme used.
 A current source attached to the left Pt strip supplies a constant current $I$. The voltage drop $V_{\mathrm{loc}}$ along the same Pt strip, as well as the non-local voltage drop $V_{\mathrm{nl}}$ along the second Pt strip, are simultaneously recorded. Panels (b), (c), and (d) show the different magnetic field rotation planes and the corresponding magnetic field orientation angles $\alpha$, $\beta$ and $\gamma$, respectively. }
 \end{figure}
In this letter, we systematically compare the magnetization-orientation dependent evolution of the (non-local) MMR and the (local) SMR in YIG/Pt nanostructures. We have simultaneously measured the MMR and SMR as sketched in Fig.~\ref{Fig:expt-schematics}, rotating the externally applied magnetic field of fixed magnitude in three mutually orthogonal planes. Our data taken close to room temperature corroborate the picture that the MMR is mediated by magnon diffusion, and reveal a qualitatively different evolution of SMR and MMR as a function of temperature.

The YIG/Pt bilayers investigated were obtained starting from a commercially available, $3\,\mu\mathrm{m}$ thick YIG film grown onto GGG via liquid phase epitaxy. The as-purchased YIG films were cleaned in a so-called Piranha etch solution (3 volumes $\mathrm{H}_2\mathrm{SO}_4$ mixed with 1 volume $\mathrm{H}_2\mathrm{O}_2$) for 120 seconds and annealed in $50\,\mathrm{\mu bar}$ oxygen for 40 minutes at $500\,^\circ \mathrm{C}$. Without breaking the vacuum, the samples were then transferred to an electron beam evaporation chamber, where we deposited a $9.6\,\mathrm{nm}$ thick Pt film. After removing the sample from the vacuum chamber, the Pt strips were defined using a combination of electron beam lithography and Argon ion beam milling. The Pt strips studied here are $100\,\mathrm{\mu m}$ long and have a lateral width of $w=1\,\mathrm{\mu m}$. We focus on a device with a strip separation (edge to edge, see Fig.\,\ref{Fig:expt-schematics}) of $d=200\,\mathrm{nm}$ in the following, but have also studied devices with $d=500\,\mathrm{nm}$ and $d=1000\,\mathrm{nm}$. For the magneto-transport experiments, the YIG/Pt nanostructures were wire-bonded to a chip carrier and inserted into the variable temperature insert of a superconducting 3D vector magnet cryostat, allowing to rotate the externally applied magnetic field $\mu_0 H\le2\,\mathrm{T}$ in any desired plane with respect to the sample. The magnetotransport data were taken by current-biasing one Pt strip with $I=100\,\mathrm{\mu A}$ using a Keithley 2400 sourcemeter, while simultaneously recording the local voltage drop $V_{\mathrm{loc}}$ (along the strip carrying the current) as well as the non-local voltage $V_{\mathrm{nl}}$ appearing along the second, nearby Pt strip using Keithley 2182 nanovoltmeters as sketched in Fig.~\ref{Fig:expt-schematics}(a). To enhance sensitivity, we use the current reversal (delta mode) method \cite{Rueffer:Diplom:2009}. We here discuss transport data taken as a function of magnetic field orientation, for fixed field magnitude $H$. To ensure full saturation of the YIG magnetization along the externally applied magnetic field, we took all data using the maximum available magnetic field $|\mu_0 H|=2\,\mathrm{T}$.  We rotated the field in three mutually orthogonal planes, as sketched in Fig.~\ref{Fig:expt-schematics}(b),(c),(d). The rotation of $\mathbf{H}$ around the direction $\mathbf{n}$ normal to the film plane, such that the magnetic field always resides within the film plane, is referred to as ip (Fig.~\ref{Fig:expt-schematics}(b)). In the oopj rotation depicted in Fig.~\ref{Fig:expt-schematics}(c), $\mathbf{H}$ is rotated around the direction $\mathbf{j}$ (along which the charge current flows), while in the oopt rotation depicted in Fig.~\ref{Fig:expt-schematics}(d), $\mathbf{H}$ is rotated around the direction $\mathbf{t}$, which is orthogonal to $\mathbf{j}$ and $\mathbf{n}$.

\begin{figure}
 \includegraphics[width=8.5cm]{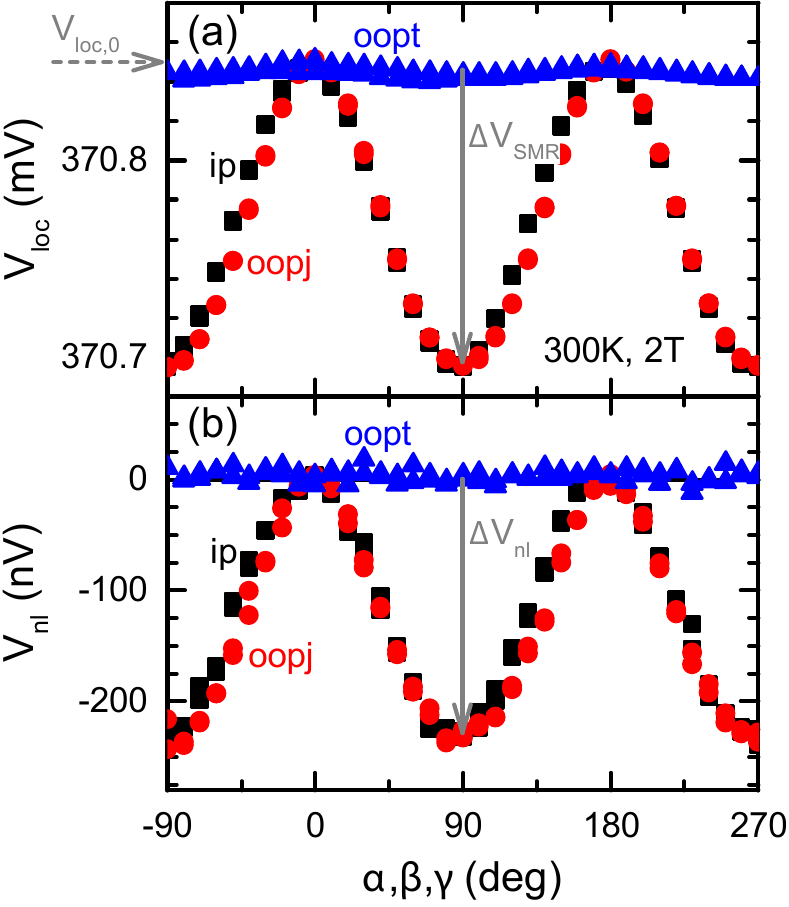}%
 \caption{\label{Fig:data} The local voltage $V_{\mathrm{loc}}$ (panel (a)) and the non-local voltage $V_{\mathrm{nl}}$ (panel (b)), recorded in two Pt strips separated by $d=200\,\mathrm{nm}$ as a function of the orientation $\alpha$ (ip), $\beta$ (oopj), $\gamma$ (oopt) of the externally applied magnetic field $\mathbf{H}$ (see Fig.~\ref{Fig:expt-schematics}). The data were taken at $T=300\,\mathrm{K}$ and $\mu_0 H=2\,\mathrm{T}$. $V_{\mathrm{loc}}$ is essentially constant in the oopt rotation plane (blue triangles), and varies in a $\sin^2$-type fashion in the ip (black rectangles) and oopj (red circles) rotation planes, respectively. The $\sin^2$ modulation with amplitude $\Delta V_{\mathrm{SMR}}<0$ (gray vertical arrow) is superimposed on a constant voltage of magnitude $V_{\mathrm{loc},0}$ (horizontal dashed arrow).  $V_{\mathrm{nl}}$ is qualitatively similar to $V_{\mathrm{loc}}$.  However, $V_{\mathrm{nl}}$ always is negative, and does not show a constant offset voltage but only a $\sin^2$-type modulation with amplitude $\Delta V_{\mathrm{nl}}$.}
\end{figure}
\begin{figure}
 \includegraphics[width=8.5cm]{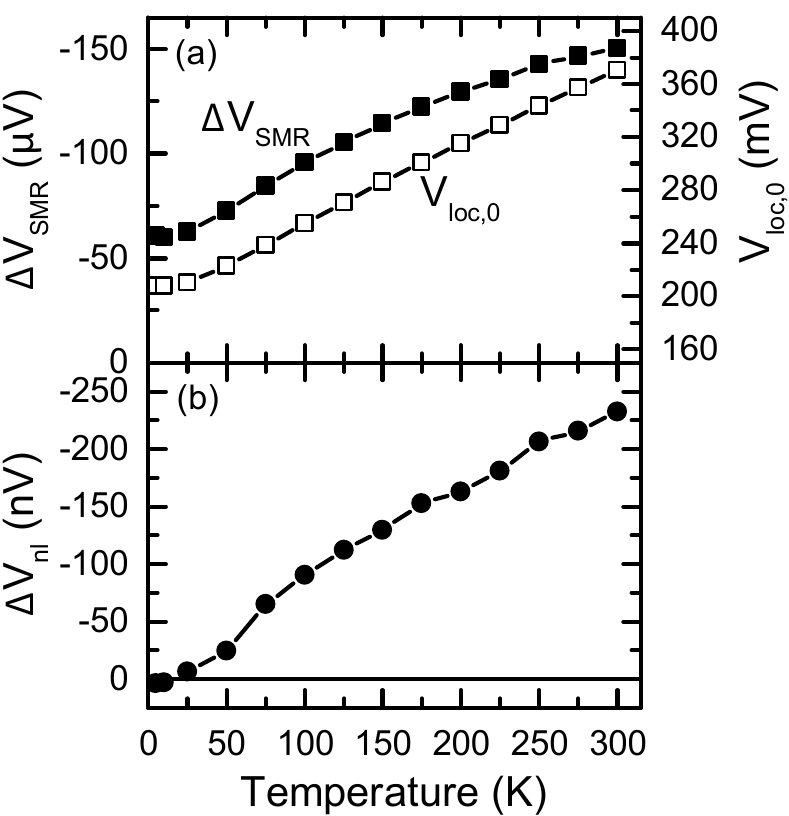}%
 \caption{\label{Fig:Tdep} Evolution of (a) the offset voltage $V_{\mathrm{loc},0}$ (open squares) and the SMR modulation voltage $\Delta V_{\mathrm{SMR}}$ (full squares) recorded in the local geometry, and of (b) the non-local voltage change $\Delta V_{\mathrm{nl}}$ (full circles) as a function of temperature for the $d=200\,\mathrm{nm}$ sample. }
 \end{figure}

Figure \ref{Fig:data} exemplarily shows magneto-transport data taken on the sample with $d=200\,\mathrm{nm}$, with the variable temperature insert thermalized to $T=300\,\mathrm{K}$. Since all data discussed in the following were taken with $|\mu_0 H|=2\,\mathrm{T}$ which exceeds the anisotropy fields in YIG by at least one order of magnitude, we assume $\mathbf{M}||\mathbf{H}$ and use the magnetic field orientations $\alpha$, $\beta$ and $\gamma$ (see Fig.\,\ref{Fig:expt-schematics}(b),(c),(d)) synonymously for $\mathbf{M}$ and $\mathbf{H}$. The local voltage $V_{\mathrm{loc}}$ depicted in Fig.\,\ref{Fig:data}(a) exhibits the dependence on magnetization orientation characteristic of the SMR. Upon rotating the external magnetic field in the plane of the YIG film (ip, see Fig.\,\ref{Fig:expt-schematics}(b)), or in the plane perpendicular to $\mathbf{j}$ (oopj, Fig.\,\ref{Fig:expt-schematics}(c)), a $\sin^2$-like modulation of $V_{\mathrm{loc}}$ with amplitude $\Delta V_{\mathrm{SMR}}<0$ on top of a constant level $V_{\mathrm{loc},0}$ is observed. $V_{\mathrm{loc},0}$ hereby is the voltage level observed when the YIG magnetization is along the $\mathbf{t}$ direction (e.g., $\alpha=90^\circ$ in ip or $\beta=90^\circ$ in oopj). The magnitude of the SMR $\left| \Delta V_{\mathrm{SMR}} \right| /V_{\mathrm{loc},0}\approx 4.5\times 10^{-4}$ agrees reasonably well with the SMR amplitude $\approx 6\times 10^{-4}$ observed in YIG/Pt heterostructures in which the $\approx 10\,\mathrm{nm}$ thick Pt films were deposited in-situ, directly after the YIG growth process \cite{Althammer:SMR:experiment:PRB:2013}. The evolution of $\Delta V_{\mathrm{SMR}}$ and $V_{\mathrm{loc},0}$ with temperature is shown in Fig.\,\ref{Fig:Tdep}(a). $\left| \Delta V_{\mathrm{SMR}}(T)\right|$ monotonically decreases by about a factor of 3 from $T=300\,\mathrm{K}$ to $T=5\,\mathrm{K}$, very similarly to the behaviour observed in other YIG/Pt samples fabricated at the Walther-Meissner-Institut \cite{Meyer:SMR:T-dep:APL:2014}. $V_{\mathrm{loc},0}$ only decreases by about 15\% in the same temperature interval, showing that defect or surface scattering is very strong in the thin Pt film.

The non-local voltage $V_{\mathrm{nl}}$ recorded in the same experiment is shown in Fig.\,\ref{Fig:data}(b). $V_{\mathrm{nl}}$ is qualitatively very similar to $V_{\mathrm{loc}}$, showing a $\sin^2$-like modulation with amplitude $\Delta V_{\mathrm{nl}}$. We would like to stress, however, that on the one hand, there is no finite constant offset in $V_{\mathrm{nl}}$, such that $V_{\mathrm{nl}}=0$ (to within the experimental noise) for the oopt rotation. On the other hand, $V_{\mathrm{nl}}$ invariably assumes {\it negative} values ($V_{\mathrm{nl}}\le 0$). According to our wiring scheme (Fig.\,\ref{Fig:expt-schematics}(a)), a negative non-local voltage implies that the non-local inverse spin Hall charge current arising in the second Pt strip (due to a diffusion of the magnon accumulation generated beneath the first Pt strip) must flow in the same direction as the charge current in the first Pt strip. Since we detect $V_{\mathrm{nl}}$ using open circuit boundary conditions, this non-local ISHE charge current is exactly balanced by an electric potential of opposite (that is negative) sign. The negative sign of $\Delta V_{\mathrm{nl}}$ in our experiments thus is consistent with the positive non-local $\Delta R>0$ reported by Cornelissen et al.~\cite{Magnon-MR:Cornelissen:arXiv2015}, since these authors use an inverted sign convention for the non-local voltage signal. The data shown in Fig.\,\ref{Fig:data} furthermore are consistent with the notion that magnon accumulation is at the origin of $V_{\mathrm{nl}}$, since one would expect maximum magnon diffusion signal (maximum $V_{\mathrm{nl}}<0$ in our experiment) for $\mathbf{M} || \mathbf{\sigma} || \mathbf{t}$ and minimal magnon diffusion signal ($V_{\mathrm{nl}}=0$) for $\mathbf{M} \perp \mathbf{\sigma}$, which translates to $V_{\mathrm{nl}}=0$ for $\mathbf{M} || \mathbf{j}$ and $\mathbf{M} || \mathbf{n}$. The non-local voltage observed in our experiment in all three rotation planes indeed confirms this expectation.
Note also that magneto-thermal (spin Seebeck) voltages cannot account for $V_{\mathrm{nl}}$, since these have a qualitatively different dependence on magnetization orientation \cite{longitudinal-spin-Seebeck:Uchida:APL:2010:172505,Magnon-MR:Cornelissen:arXiv2015}.

The magnitude $\left| \Delta V_{\mathrm{nl}} \right| \approx 250\,\mathrm{nV}$ of the magnetization-orientation dependent modulation in the non-local voltage at $T=300\,\mathrm{K}$ is about 1000 times smaller than the local $\left| \Delta V_{\mathrm{SMR}} \right|\approx 150\,\mathrm{\mu V}$. Fitting the $\Delta V_{\mathrm{nl}}$ observed for pairs of strips with separation $d=200\,\mathrm{nm}$, $d=500\,\mathrm{nm}$ and $d=1\,\mathrm{\mu m}$, respectively, using $\Delta V_{\mathrm{nl}}=(C/\lambda) \exp(d/\lambda)/(1- \exp(2 d/\lambda))$ derived as Eq.\,(7) in Ref.~\cite{Magnon-MR:Cornelissen:arXiv2015} for 1D spin diffusion, we obtain $\lambda \approx 700\,\mathrm{nm}$. This value of $\lambda$ is about one order of magnitude smaller than the value reported by Cornelissen et al.~\cite{Magnon-MR:Cornelissen:arXiv2015} for their samples. The discrepancy might be evidence for enhanced magnon scattering owing to YIG surface damage caused by our fabrication process. In addition, $\lambda \approx 700\,\mathrm{nm}$ is smaller than the YIG film thickness of $3\,\mathrm{\mu m}$ in our case, suggesting that diffusion in more than one dimension could be important. To conclusively resolve this point, multiple samples with different YIG film thicknesses and a series of different Pt strip separations $d$ must be systematically compared, which is beyond the scope of this work.

Interestingly, the temperature dependencies of $\Delta V_{\mathrm{nl}}$ and $\Delta V_{\mathrm{SMR}}$ are very different. As evident from Fig.\,\ref{Fig:Tdep}(b), the magnitude of $\Delta V_{\mathrm{SMR}}$ at low $T$ is only about a factor of 3 smaller than at room temperature, while $\Delta V_{\mathrm{nl}}=0$ for $T\le10\,\mathrm{K}$. The strong decrease in $\Delta V_{\mathrm{nl}}(T)$ can be rationalized considering an increase of the magnon propagation length $\lambda$ with decreasing $T$. In a simple picture, the non-equilibrium magnons generated at the YIG/Pt interface spread across a volume $V_{\mathrm{mag}}\propto \lambda^3$, such that the magnon accumulation (viz.~the non-equilibrium magnon density) scaling with $1/V_{\mathrm{mag}}$ decreases with $T$. In the limit of infinite $\lambda$, the magnon accumulation and thus also $\Delta V_{\mathrm{nl}}$ vanishes. More sophisticated theoretical analyses corroborate this intuitive picture \cite{YIG-Pt:Xiao:arXiv2015,YIG-Pt:Bender:PhysRevB.91.140402:2015}. Note also that the finite SMR signal at low $T$ is direct evidence that the spin Hall effect is only weakly temperature dependent \cite{Meyer:SMR:T-dep:APL:2014}, such that the decrease of the MMR (viz.~of $\Delta V_{\mathrm{nl}}$) with $T$ cannot be simply attributed to spin Hall physics.

In conclusion, we have simultaneously measured the local and the non-local magnetoresistive response of two parallel Pt strips separated by a gap of a few 100\,nm, deposited onto yttrium iron garnet. The local magnetoresistance (current-biasing one Pt strip and measuring the magnetization-orientation dependent voltage drop along this same Pt strip) shows the characteristic fingerprint of spin Hall magnetoresistance, as expected for a YIG/Pt heterostructure. We furthermore observe a non-local voltage $V_{\mathrm{nl}}$ along the second, electrically isolated Pt strip upon current biasing the first one. Our data taken at room temperature confirm the results put forward by Cornelissen et al. \cite{Magnon-MR:Cornelissen:arXiv2015}. In addition, we have measured $V_{\mathrm{nl}}$ as a function of magnetization orientation in three mutually orthogonal rotation planes, and studied the evolution of both the local and the non-local magnetoresistance from room temperature down to $5\,\mathrm{K}$. All our experimental data can be consistently understood assuming that the non-local magnetoresistance is mediated via magnon accumulation.

We gratefully acknowledge discussions with G.\,E.\,W.\,Bauer and S.\,Klingler, and funding via the priority programme spin-caloric transport (spinCAT) of Deutsche Forschungsgemeinschaft, project GO 944/4.

%\begin{acknowledgments}
%\end{acknowledgments}

%\bibliography{aipsamp}
%\bibliography{bib-APL-nonlocal-MR}

\end{document}